\documentclass[fleqn,usenatbib]{mnras}

\usepackage{newtxtext,newtxmath}

\usepackage[T1]{fontenc}
\usepackage{ae,aecompl}

\usepackage{graphicx}
\usepackage{amsmath}
\usepackage{amssymb}
\usepackage{aas_macros}

\newcommand{\kpc}{~\mathrm{kpc}}
\newcommand{\Mpc}{~\mathrm{Mpc}}
\newcommand{\dex}{~\mathrm{dex}}
\newcommand{\Mo}{~\mathrm{M}_\odot}
\newcommand{\Mokpcmt}{~\mathrm{M}_\odot\,\mathrm{kpc}^{-2}}
\newcommand{\arcsectwo}{~\mathrm{arcsec}^2}
\newcommand{\magarcsecmtwo}{~\mathrm{mag}\,\mathrm{arcsec}^{-2}}
\newcommand{\mtol}{~\mathrm{M}_\odot\,\mathrm{L}^{-1}_\odot}
\newcommand{\kmsMpc}{~\mathrm{km}\,\mathrm{Mpc}^{-1}\,\mathrm{s}^{-1}}

\title[ICL as a tracer of the matter density distribution]{The intra-cluster light as a tracer of the total matter density distribution: a view from simulations}

\author[I. Alonso Asensio et al.]
{Isaac Alonso Asensio,$^{1,2}$\thanks{E-mail: isaacaa@iac.es (IAA)}
Claudio Dalla Vecchia,$^{1,2}$
Yannick M.~Bah\'e,$^{3}$\newauthor
David J.~Barnes$^{4}$ and
Scott T.~Kay$^{5}$
\\
$^1$Instituto de Astrof\'isica de Canarias, C/V\'ia L\'actea s/n, E-38205 La Laguna, Tenerife, Spain  \\
$^2$Departamento de Astrof\'isica, Universidad de La Laguna, Av.~Astrof\'isico Francisco S\'anchez s/n, E-38206 La Laguna, Tenerife, Spain \\
$^3$Leiden Observatory, Leiden University, PO Box 9513, 2300 RA Leiden, The Netherlands  \\
$^4$Department of Physics, Kavli Institute for Astrophysics and Space Research, Massachusetts Institute of Technology,\\\hspace{4pt} Cambridge, MA 02139, USA  \\
$^5$Jodrell Bank Centre for Astrophysics, Department of Physics and Astronomy, School of Natural Sciences,\\\hspace{4pt} The University of Manchester, Manchester M13 9PL, UK
}

\date{Accepted XXX. Received YYY; in original form ZZZ}

\pubyear{2020}

\begin{document}
\label{firstpage}
\pagerange{\pageref{firstpage}--\pageref{lastpage}}
\maketitle

\begin{abstract}
   By using deep observations of clusters of galaxies, it has been recently found that the projected stellar mass density closely follows the projected total (dark and baryonic) mass density within the innermost $\sim 140$ kpc. 
   In this work, we aim to test these observations using the Cluster-EAGLE simulations, comparing the projected densities inferred directly from the simulations. 
   We compare the iso-density contours using the procedure of Montes \& Trujillo (2019), and find that the shape of the stellar mass distribution follows that of the total matter even more closely than observed, although their radial profiles differ substantially. 
   The ratio between stellar and total matter density profiles in circular apertures, shows a slope close to $-1$, with a small dependence on the cluster's total mass. 
   We propose an indirect method to calculate the halo mass and mass density profile from the radial profile of the intra-cluster stellar mass density.
\end{abstract}

\begin{keywords}
galaxies: clusters: general -- methods: numerical
\end{keywords}

\section{Introduction} \label{sec:intro}

Ten to more than thirty percent of the stellar light of clusters of galaxies comes from a diffuse distribution of stars emitting the so called intra-cluster light (ICL), the inferred fraction depending on the definition of the border between the brightest central galaxy and the diffuse stellar component, the radial extent at which the stellar mass distribution is integrated and the relaxation state of the clusters \citep[e.g.,][]{Krick07,Gonzalez2013,Mihos2017,JimenezTeja18,Zhang19}. This distribution is produced by the stripping of stars from galaxies undergoing mergers and tidal interactions during their evolution in the cluster environment \citep[see][for a review]{Mihos15}. 
Due to its low surface brightness, the observational study of the stellar population producing the intra-cluster light has been challenging. 
An increasing effort towards deep imaging of clusters of galaxies in the recent years, both through individual cluster imaging, up to $z\simeq 1.5$, \citep[e.g.,][]{Mihos05,Montes2014,Burke15,Morishita2017,Ko2018,JimenezTeja18,Montes18,DeMaio2018,DeMaio2020}, and by stacking observations of multiple clusters \citep[e.g.,][]{Zibetti2005,Zhang19} has allowed new insights into the ICL.

\begin{figure*}
\includegraphics[width=0.95\textwidth]{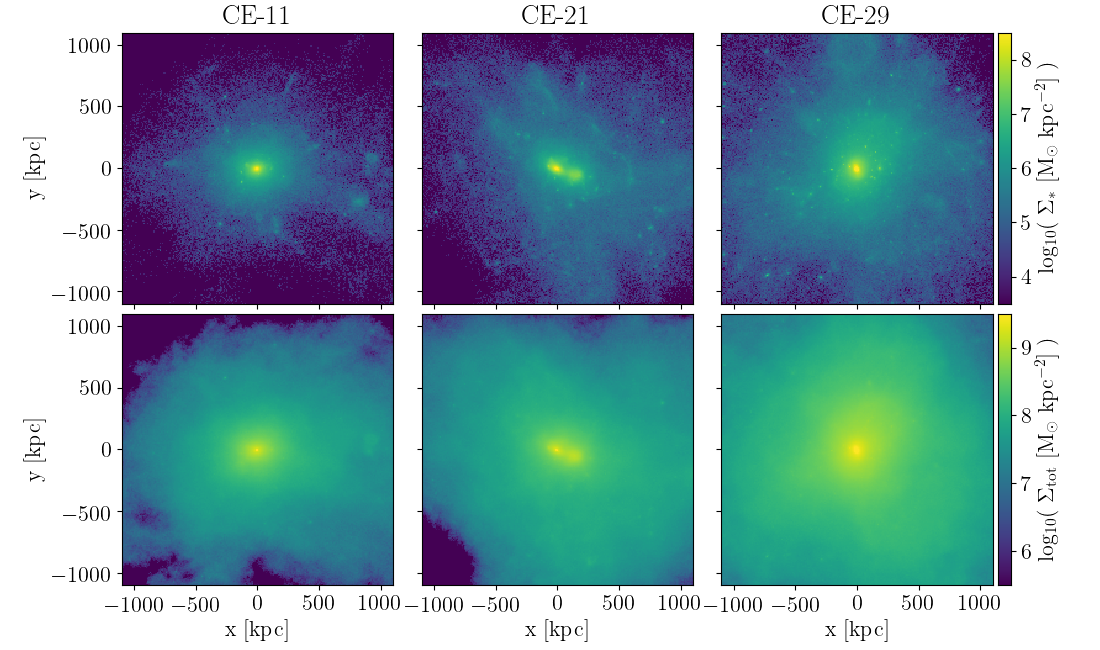}
\caption{Projected density of stars (top) and matter (bottom) for three different cluster of the C-EAGLE simulations at $z=0.352$. The secondary density peak in the Cluster CE-21 (middle panel) can be due to a recent major merger event which has stripped stars from the interacting galaxies, or numerical artifacts produced by \texttt{SUBFIND} not being able to correctly assign stellar particles to substructures.
\label{fig:projdensity}}
\end{figure*}

One recent, remarkable result achieved with deep imaging is the tight correlation between the distribution of the stellar surface density, inferred from its surface brightness, and the surface density of the total mass, measured by modelling the gravitational lensing signal \citep[][hereafter MT19]{Montes19}. MT19 proposed that the surface density of the stellar mass not bound to galaxies should settle in the potential well of the cluster similarly to the dark matter. This could be used to trace the total matter distribution of clusters within a cluster-centric distance set by the depth of the observations. They also compared their result with total mass surface densities inferred from the X-ray emission of the intra-cluster medium, and concluded that this method is limited by the misalignment of the gaseous component with respect to the dark matter and stellar mass in non-relaxed clusters. Their quantitative analysis made use of the Modified Haussdorf Distance (MHD) \citep{Dubuisson94} to quantify the deviation between iso-density contours of stars and total matter. They found that, in general, the stellar surface density has smaller MHD values than that of the intra-cluster medium where both are compared with the iso-density contours of total mass.

In this Letter, we test this observational result with state-of-the-art cosmological, hydrodynamic simulations of the Cluster-EAGLE project \citep[C-EAGLE, ][]{Barnes17,Bahe17}. We give a brief description of the simulations in the next section and a description of the analysis in section~\ref{sec:shapematching}. The main results of this work are shown in section~\ref{sec:results} and discussed in section~\ref{sec:discussion}, along with some concluding remarks.

\section{Simulations} \label{sec:simulations}

We have used the set of 30 zoom-in cluster simulations performed within the C-EAGLE project. The simulated clusters are uniformly distributed in the mass range $10^{14} < M_{200}/\mathrm{M}_\odot < 10^{15.4}$, where $M_{200}$ is the halo mass.\footnote{$M_{200}$ is the mass enclosed in a sphere of radius $r_{200}$, whose mean density equals 200 times the critical density of the Universe.} The simulations were performed with the EAGLE model for galaxy formation and evolution, with the AGNdT9 calibration \citep{Schaye15}. 
They provide a physical spatial resolution of $\epsilon=0.7\kpc$ (at $z<2.8$) and baryonic mass resolution of $m_{\mathrm{gas}} \approx 1.81 \times 10^{6}\Mo$. For more information on the EAGLE model and its comparison with global relations of the observed galaxy population, the reader is referred to \cite{Schaye15} and \cite{Crain2015}. 

For more details on the numerical algorithms describing photo-ionization equilibrium cooling, star formation, stellar evolution, stellar feedback, black hole growth and feedback, and the hydrodynamic scheme we refer the reader to \cite{Wiersma2009a}, \cite{Schaye2008}, \cite{Wiersma2009b}, \cite{DallaVecchia2012}, \cite{Rosas-Guevara2015}, and \cite{Schaller2015}, respectively.

\begin{figure}
\centering
\includegraphics[width=0.45\textwidth]{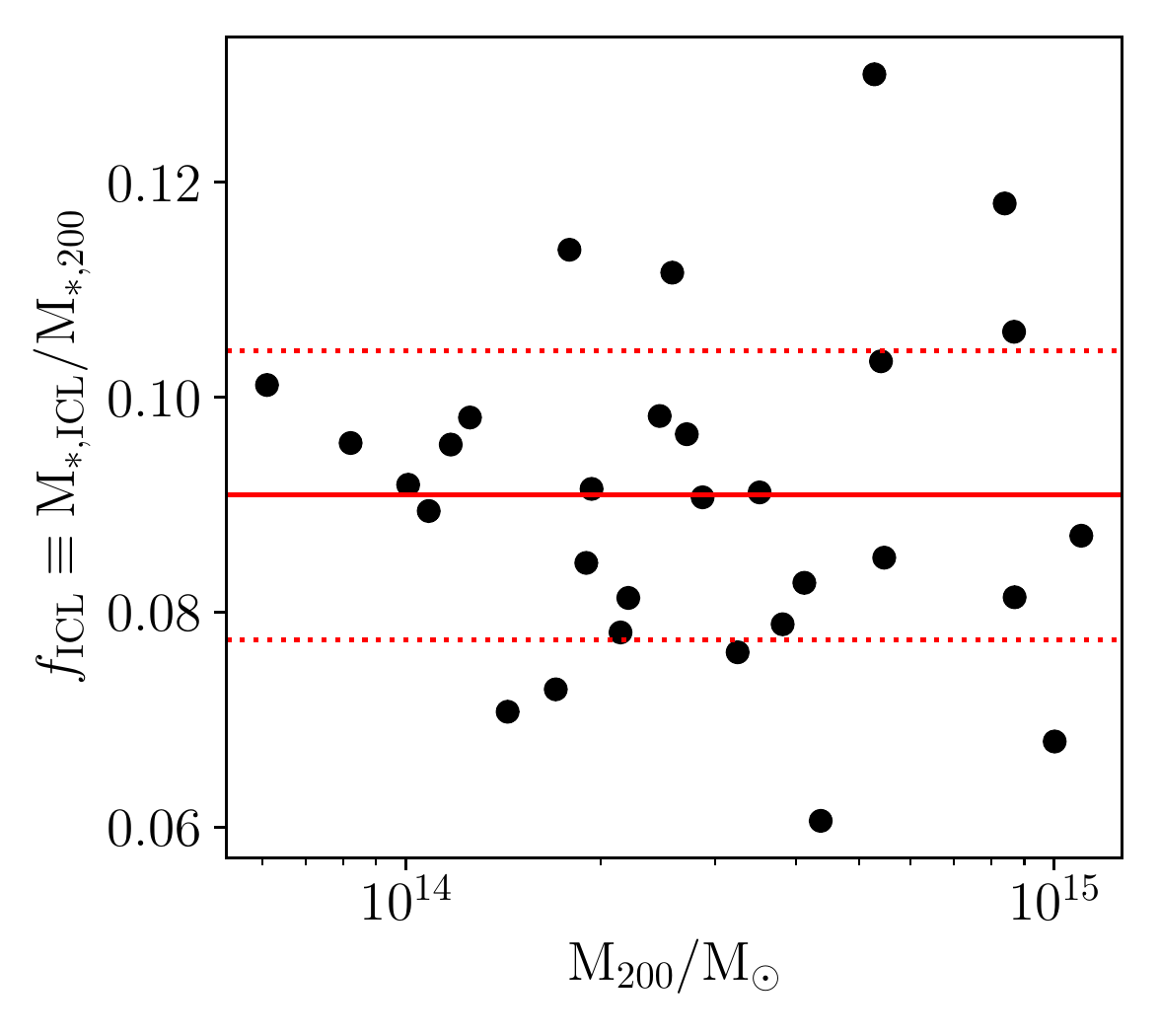}
\caption{Fraction of stellar mass contributing to the ICL, $f_\mathrm{ICL}$, as function of halo mass, $M_{200}$, for all clusters in the sample. There is no evidence for a correlation with halo mass. The solid line marks the average value of $f_\mathrm{ICL}$, and the dashed lines the spread around it.}
\label{fig:icl_frac}
\end{figure}

For the results presented here, we have used the particle data, friends-of-friends and \texttt{SUBFIND} \citep{Dolag2009} groups at $z=0.352$ to match the average redshift of the Hubble Frontier-Fields clusters \citep{Lotz2017}. Furthermore, the same analysis was performed at $z=0$, and we found no significant difference.
Throughout the paper we assume the cosmological parameters of the C-EAGLE simulations, $(\Omega_0,\Omega_\Lambda,h,n_\mathrm{s},\sigma_8)=(0.307,0.693,0.6777,0.961,0.8288)$ \citep{Planck2014}, where $\Omega_0$ and $\Omega_\Lambda$ are the matter and dark energy fractions, $h$ is the Hubble constant in units of $100\kmsMpc$, $n_\mathrm{s}$ and $\sigma_8$ are the spectral index and the power spectrum normalisation used to generate the initial conditions.

In the analysis, we have used all particles belonging to the main halo of the largest friends-of-friends group in each simulation, i.e., we excluded all particles bound to satellite galaxies and substructures within the same friends-of-friends group. Maps of projected stellar and total matter density were produced with a spatial resolution of $5\kpc$, in order to mimic the spatial resolution employed in the analysis of the observational data ($3\times 3\arcsectwo$ at $z\simeq 0.35$). We have repeated the analysis with higher ($3.75\kpc$) and lower ($7.5\kpc$) resolution without finding any remarkable difference.
The main advantage with respect to observations is that there is no need of masking the light of satellite galaxies. 
However, debris from tidal interactions between galaxies will be included in the projected matter density.
Furthermore, there are biases due to \texttt{SUBFIND} failing to assign stellar particles to satellites (Bah\'e et al., in prep).

\begin{figure*}
\includegraphics[width=0.95\textwidth]{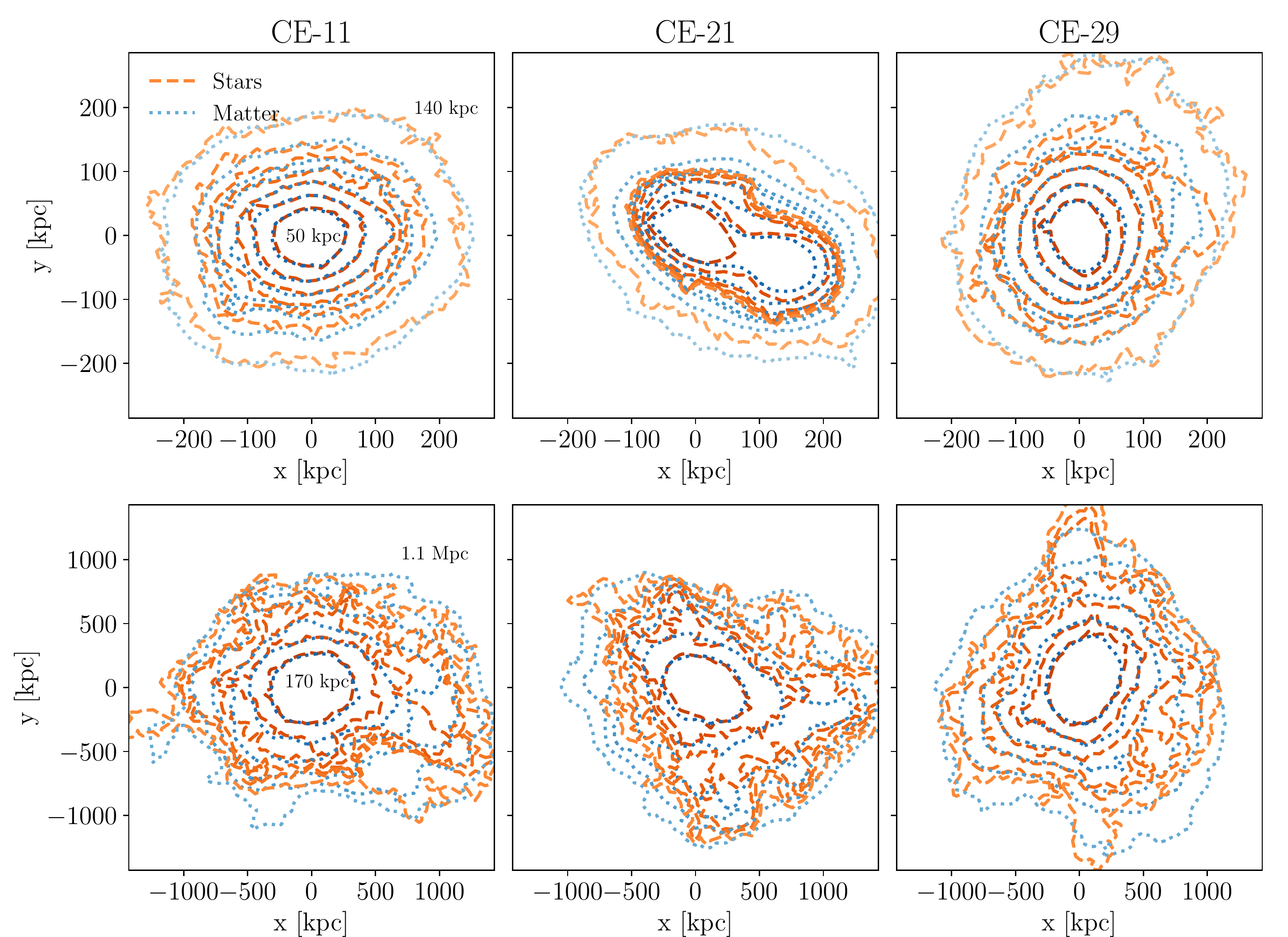}
\caption{Isodensity contours of the inner ($R\leq 140\kpc$) and outer ($R>140\kpc$) regions (top and bottom, respectively) of total matter (blue dotted lines) and stars (red dashed lines) for three different clusters. Lighter colours indicate larger distances (lower densities) from the centre.\label{fig:contour}}
\end{figure*}

Examples of the projected stellar and total mass density are shown in Fig.~\ref{fig:projdensity} for three simulated clusters of increasing virial mass. 
The top row corresponds to the projected density of stars, while the bottom row shows the density of total matter (dark and baryonic). 

Uncertainties on the amount of ICL mass produced and its radial distribution may arise from the modelling of the star formation rate and the spatial and mass resolution of numerical simulations. The EAGLE model matches quite accurately the observed stellar mass and luminosity functions \citep{Schaye15,Trayford2015}. Moreover, it reproduces the evolution of the stellar mass function and the observationally inferred density of stars in the universe up to high redshift ($z=7$) \citep{Furlong2015}. However, while the reference simulation matches the observed sizes of galaxies over several decades in stellar mass, the AGNdT9 calibration yields an offset in the relation towards more compact galaxies \citep{Schaye15}. This last point seems to be relevant in the interpretation of the ICL mass fractions described in the next section, where the inferred values are on the low side of the distribution of those derived from observations (see references in section~\ref{sec:intro}): compact galaxies are less prone to stripping. On the other hand, \cite{Henden2019} noted that having too large in size galaxies in their simulations boosts the effect of tidal stripping, increasing the fraction of stellar mass in the ICL, and that uncertainties in galaxy sizes are the major contributors to the uncertainty in the determining the fraction of mass in the ICL in simulations.

\section{Analysis} \label{sec:shapematching}

Before describing the methodology used in the analysis of the simulation data, we briefly discuss a consistency check for the simulated clusters. We computed the fraction of stellar mass in the ICL, $f_\mathrm{ICL}$, and compared it with expected observational and theoretical values.
For the sake of ease, we adopted the methodology of \citep{Rudick11}.
The mass fraction has been computed as the stellar mass with projected stellar density below some threshold surface brightness, $\mu$, with respect to the total stellar mass within $r_{200}$.
As in \citep{Rudick11}, we have converted the stellar surface density into surface brightness assuming a constant mass-to-light ratio of $5\mtol$, and set $\mu = 26.5\magarcsecmtwo$ as the threshold.

We show in figure~\ref{fig:icl_frac} the computed $f_\mathrm{ICL}$ as function of halo mass. We find that $f_\mathrm{ICL}=0.091\pm 0.013$ (solid and dashed lines), with no significant correlation with the total mass of the clusters (the Pearson correlation coefficient is $0.0063$). The result is consistent with that of \citep{Rudick11}. Although the range of halo masses in our sample is rather narrower, similar fractions and the lack of correlation have been reported by \citep{Pillepich2018}, when using a definition of the ICL related to the size of the central galaxy.  The result is consistent with previous simulations \citep{Rudick11,Contini14}, where they applied semi-analytical models to N-body simulations, and hydrodynamical cosmological simulations \citep{Pillepich2018,Henden2019}. 
Finally, observations using similar thresholds have reported as well similar mass fractions \citep{Krick07,Montes2014}.

We have followed a methodology similar to MT19 to extract iso-density contours. We computed circularly averaged radial profiles of the density of the stellar and total mass. For this, we take the position of the minimum of the potential energy as centre of the cluster \citep{McAlpine16}. The projected densities for drawing the contours\footnote{We used the \texttt{contour} function of \texttt{matplotlib} to compute the contours.} were selected interpolating the profiles at radii of 50, 75, 100, 125, $140\kpc$ (the distances used by MT19) for the inner part, and of 170, 220, 300, 460, 620, 780, 940, $1100\kpc$ for the outer regions, and only up to $r_{200}$. 
At large distance from the centre of the clusters ($r>140\kpc$), we down-sample the images merging $4\times 4$ pixels, thus degrading the spatial resolution to $20\kpc$, to smooth the otherwise very noisy contours. 
The contours of the projected densities are shown in Fig.~\ref{fig:contour}, for the same three clusters as depicted in Fig.~\ref{fig:projdensity}. The projected total mass density contours are drawn with blue dotted lines, and the projected stellar density contours with red dashed lines, where a darker colour indicates a smaller radius. 
The top row is a close-up view of the contours near the centre of the clusters, out to $140\kpc$, whereas the contours at larger distances are shown in the bottom row.

We measured projected radial distances from the centre of the cluster instead of elliptical distances to the centre of the brightest central galaxy, as usually done in observations. 
This simplification is not crucial to derive the iso-density contours, as it only changes the values of density at which the contours will be drawn. 
In practice, this means that the distances we use are systematically different from those of MT19, the difference depending on the eccentricity of the brightest central galaxy, or the presence of more than one central galaxy, that we excluded from the analysis, or both. As this is only an exploratory analysis we ignore these differences.

As in MT19, to compare the shape of the contours, we estimated the Modified Hausdorff distance (MHD) defined by \cite{Dubuisson94}:
\begin{equation}
d_{\mathrm{MH}}(X, Y) = \max \left( d(X, Y) , d(Y,X)\right) , \\
\end{equation}
where
\begin{equation}
d(X,Y) = \frac{1}{N_X} \sum_{\textbf{x} \in X} \min_{\textbf{y} \in Y} \Vert \textbf{x} - \textbf{y} \Vert.
\end{equation}
The two samples, $X \equiv \{\textbf{x}_1, \textbf{x}_2, \dots, \textbf{x}_{N_x}\}$ and $Y \equiv \{ \textbf{y}_1, \textbf{y}_2, \dots, \textbf{y}_{N_y} \}$, contains the points defining two contours, and $\Vert \cdot \Vert$ is the Euclidean norm. As we may have different closed contours for the same density value, we select for each distance the contour composed by the largest number of segments. The selected contours are shown in Fig.~\ref{fig:contour}.

\begin{figure*}
\centering
\includegraphics[width=0.45\textwidth]{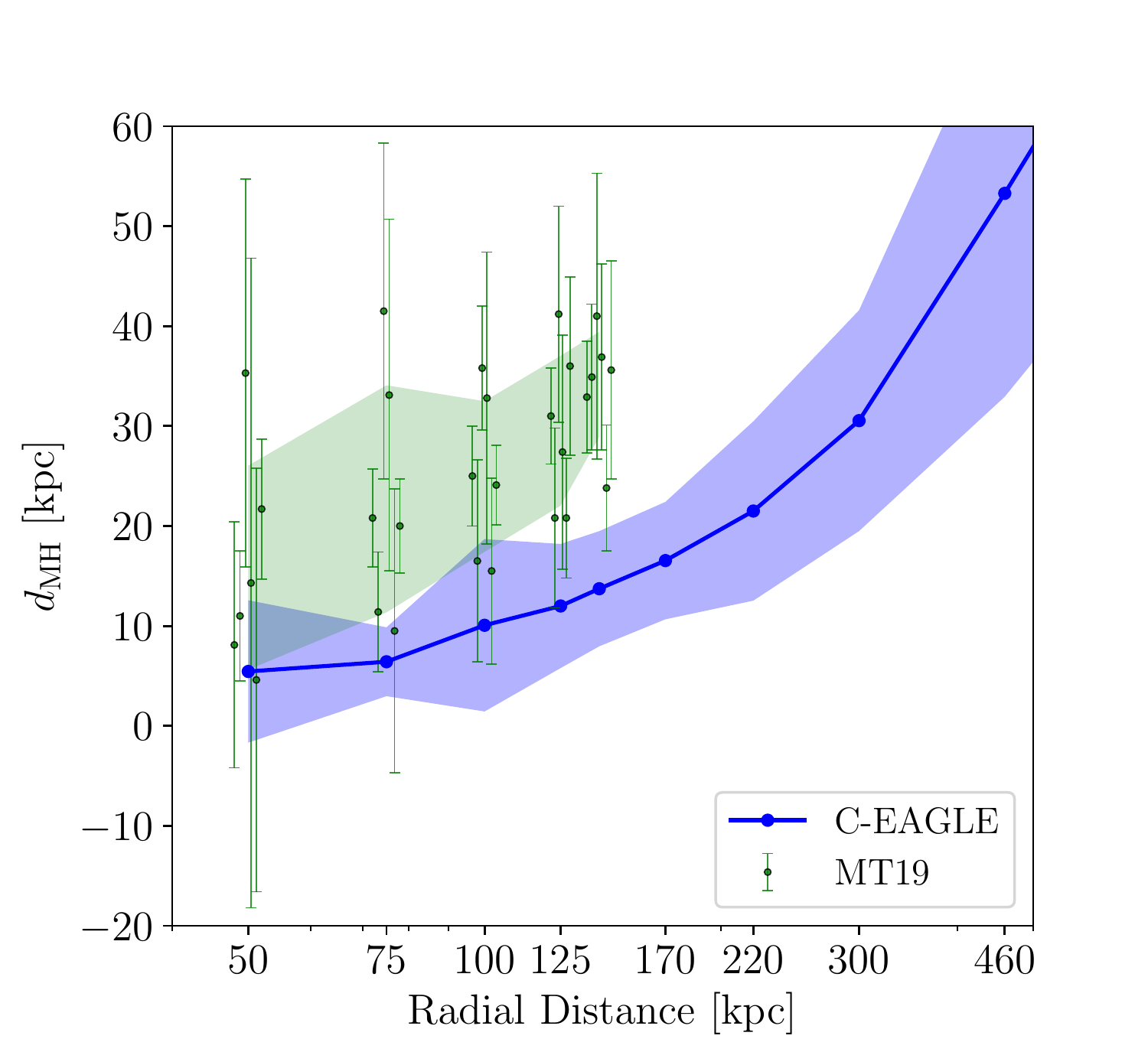}
\includegraphics[width=0.45\textwidth]{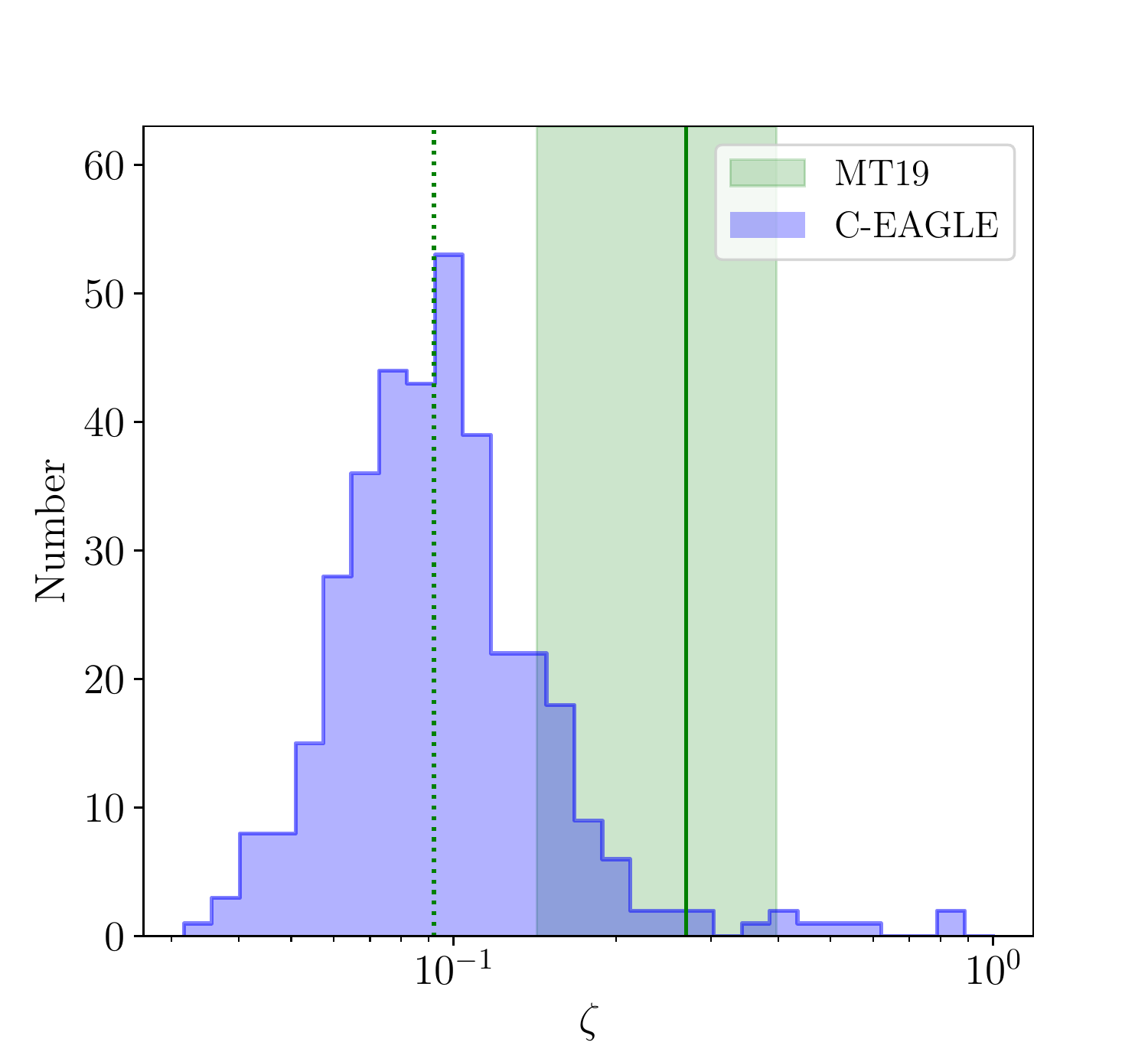}
\caption{\textit{Left panel.} Comparison of the MHD of MT19 (in green, signle measurements with error bars) and that from the C-EAGLE simulations (in blue, solid line). The shadows indicate the 1-$\sigma$ region for each method. A small scatter in the radial distance of the MT19 data has been added for clarity.
\textit{Right panel.} Histogram of $\zeta$ computed from all the contours taken inside the virial radius of each C-EAGLE cluster. The vertical (green) solid line represents the mean value of $\zeta$ obtained by MT19, embedded in its 1-$\sigma$ region. The dotted, vertical line indicates their lowest value.}
\label{fig:mhd_local}
\end{figure*}

When measuring the MHD close to the virial radius of the clusters, we would expect an increase of its value, as the outskirts of clusters are not dynamically relaxed and fewer stellar particles are populating it, producing noisier contours. In order to compare the MHD across different distances, we define the relative MHD as
\begin{equation}
\zeta = \frac{d_\mathrm{MH}(r)}{r}\,,
\end{equation}
where $r$ is the distance at which the iso-density contours have been computed. This way, we are measuring deviations as fraction of the distance. 
We find that this definition removes almost entirely the correlation with distance.

\section{Results} \label{sec:results}

In the left panel of Fig.~\ref{fig:mhd_local}, we show with the blue, solid line the mean value of $d_\mathrm{MH}$, the shaded area depicts the 1-$\sigma$ confidence interval.
We overplotted the MHDs calculated by MT19, as well as their 1-$\sigma$ area, in green.
For sake of clarity, observational points for individual cluster are slightly displaced along the x-axis. 
From that panel, we can highlight that:
\begin{enumerate}
   \item the $d_\mathrm{MH}$ from both simulations and observations are of the same order of magnitude;
   \item they show the same trends with radius;
   \item and simulations have a $\sim 50\%$ lower $d_\mathrm{MH}$ than observations, with smaller scatter. 
\end{enumerate}

As $d_\mathrm{MH}$ increases monotonically with the distance at which it is computed, we introduced the relative MHD, $\zeta$, to obtain a distance-free similarity measurement. We show in Fig.~\ref{fig:mhd_local} (right panel) the distribution of $\zeta$ for all contours and clusters, in blue, and the $\zeta$ extracted from MT19's data, in green. Most of the values of $\zeta$ are lower than those observed: 96 percent of the relative MHDs are below the mean observed value. The shape of the distribution is remarkably close to a Gaussian distribution in logarithmic space, with mean $\langle\zeta\rangle=0.107$ and dispersion $\sigma_\zeta=0.080$, indicating that $\zeta$ is a solid, scale-free estimate of the similarity of contours at any cluster-centric distance.

\begin{figure*}
\centering
\includegraphics[width=0.45\textwidth]{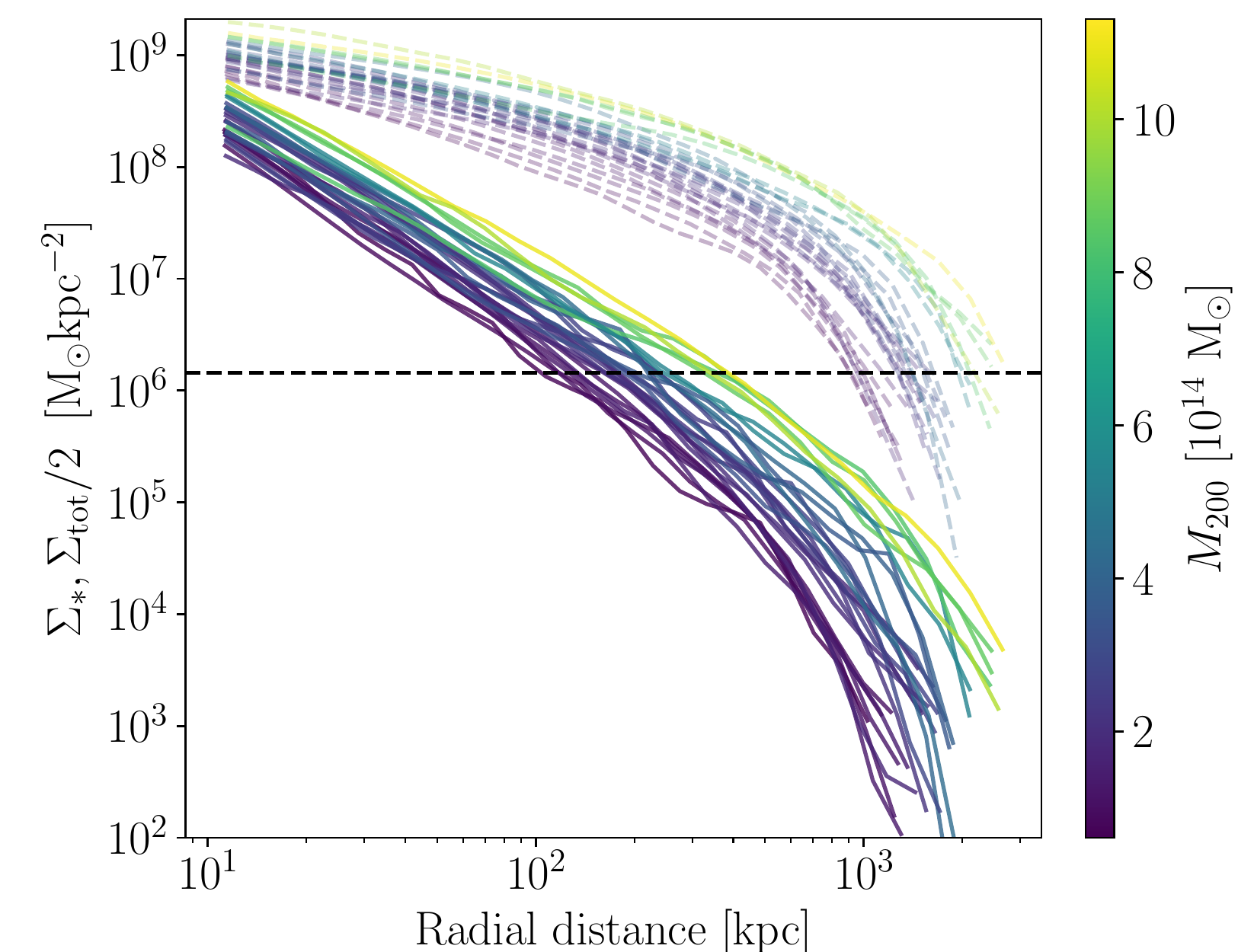}
\includegraphics[width=0.45\textwidth]{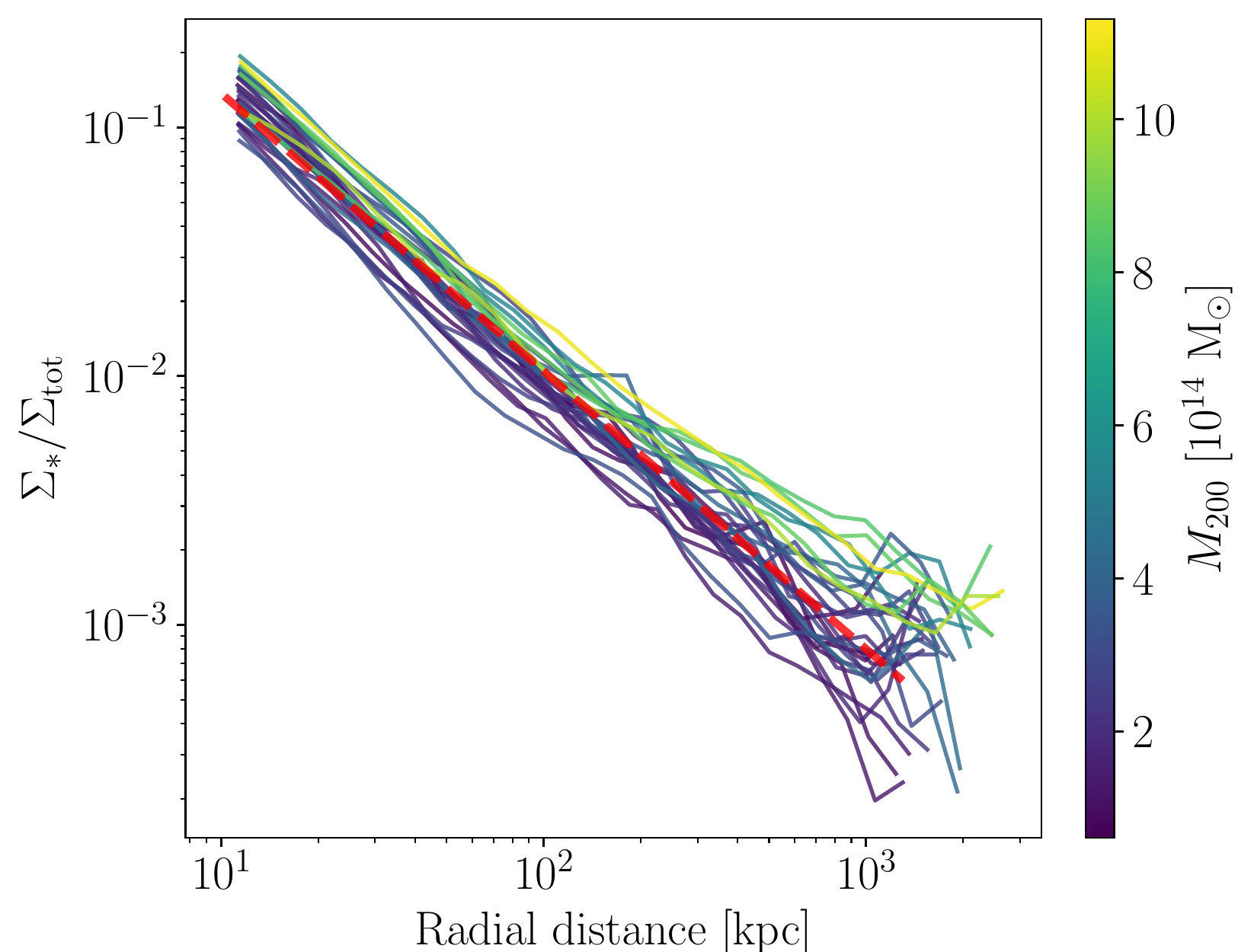}
\caption{\textit{Left panel.} Stellar (solid lines) and total matter(dashed lines) surface density profiles from the particles of the main halo of the cluster. We consider only the ICL mass (see text), including the particles not bounded to any substructure. 
The dashed line is the threshold used for computing the ICL mass fraction (i.e., $\mu = 26.5$ mag arcsec$^{-2}$, or $\Sigma_* \approx 1.4 \times 10^6 \Mokpcmt$).
\textit{Right panel.} The ratio between the stellar and total matter density profiles for all the clusters. The red, dashed line is the best-fit power law given in equation~\ref{eq:fit_ratio}. \label{fig:stellar_density}}
\end{figure*}

We would like to highlight two relevant issues with the definition of $d_\mathrm{MH}$ that can bias the observational values towards higher values. 
The $d_\mathrm{MH}$ is defined based on points and not continuous segments. 
This obviously simplifies the computation, but it has to be taken into account when dealing with coarse datasets, as two similar shapes can have a non-negligible  $d_\mathrm{MH}$.
Second, each point's contribution is defined positive and with respect to the other set of points.
This provides a distance that increases monotonically with noise \citep{Dubuisson94}, thus special care must be taken when dealing with data with low signal-to-noise or large uncertainties.
Both these points could be driving the observed $d_\mathrm{MH}$ towards higher values, as masking galaxies introduces non-continuous contours and the spatial resolution of the lensing models is limited.

In addition to the study of the similarity between the total matter and stellar mass distribution, we have also compared the density profiles of the stellar component.
In Fig.~\ref{fig:stellar_density} (left panel) we show circularly averaged density profiles of the stellar particles.
They follow a power-law behaviour up to $\sim 500\kpc$ for the lightest halos, and $\sim 1\Mpc$ for the more massive ones. 
At such distances, the interactions between substructures are weaker, and fewer particles get ejected to the intra-cluster medium, thus they can no longer successfully trace the potential well.
In the right panel of Fig.~\ref{fig:stellar_density} we show the ratio between the stellar and total matter density profiles.
This ratio is close to a power law with scatter of $0.1\dex$ and a slope of about $-1$.
We have performed a fit to all the profiles at once, with and without normalising the radial distance using $r_{200}$, yielding the relations:
\begin{align}
   \log_{10} \Sigma_\mathrm{tot} =& \log_{10} \Sigma_{*} + \nonumber\\ &(1.115 \pm 0.005) \log_{10} r - (0.25 \pm 0.01)\,, \label{eq:fit_ratio} \\
   \log_{10} \Sigma_\mathrm{tot} =& \log_{10} \Sigma_{*} + \nonumber\\ &(1.085 \pm 0.004) \log_{10} (r/r_{200}) + (3.144 \pm 0.005)\,.
   \label{eq:fit_ratio_r200}
\end{align}
The residuals of both fits have a similar scatter: $0.147$ and $0.127$~dex for equations~\ref{eq:fit_ratio} and \ref{eq:fit_ratio_r200}, respectively. We recall that the AGNdT9 feedback calibration, used in the C-EAGLE simulations, yields more compact galaxies than the reference model for stellar masses $M_\star>10^{10}\Mo$. The less efficient tidal stripping may therefore deposit more stellar mass closer to the centre of the cluster, resulting in a steeper density profile. However, this bias may be of secondary importance, at least within the central $100\kpc$ (Bah\'e et al., in prep.).

We propose a new, indirect way of measuring a cluster's mass knowing its stellar density profile in the innermost region.
First, via deep imaging as that performed by MT19, the stellar density profile can be obtained and extrapolated up to $r_{200}$ assuming a power law.
Then, using equation~\ref{eq:fit_ratio} or ~\ref{eq:fit_ratio_r200}, the total mass density profile can be computed.
This profile can be integrated to obtain an estimation of the cluster's total mass.

This procedure would be similar to that proposed by \citet{Pillepich2018}.
In that case, however, only the power law slope of the 3D stellar mass density profile was used to infer the total halo mass, in our case we use more information (the 2D stellar density profile and equation~\ref{eq:fit_ratio} or \ref{eq:fit_ratio_r200}), expecting less scatter in the mass estimate.

\section{Discussion \& Conclusions} \label{sec:discussion}

We have studied the similarity of the projected stellar and total matter distributions in the halos of massive galaxy clusters using the C-EAGLE set of 30 zoom-in simulations of clusters of galaxies.
In the analysis, we considered as constituents of the diffuse distribution of stellar mass only particles in the friends-of-friends group that were not assigned to any substructure by the \texttt{SUBFIND} algorithm.

We can summarise our results as follows:
\begin{enumerate}
   \item we confirm the finding of MT19: the projected distribution of stars closely follow the projected distribution of the total mass, although their radial profiles differ substantially;
   \item the ICL, approximated as those stars in the region where $\mu>26.5\magarcsecmtwo$ ($\Sigma_{*} \approx 1.4 \times 10^6 \Mo$), accounts for $\sim 10$ percent of the stellar content of the cluster within $r_{200}$; this fraction does not show any correlation with the mass of the cluster;
   \item the ratio between the surface density profiles of the stellar to the total matter follows a simple power-law up to the virial radius, equations~\ref{eq:fit_ratio} and~\ref{eq:fit_ratio_r200}; as the slope and amplitude of the stellar surface density profile can be extracted from observation, we proposed a method to estimate the total mass surface density profile, thus the mass of the halo;
   \item the similarity between the stellar and total matter distributions in the cluster halo is even higher in the simulations than that observed by MT19 (Fig.~\ref{fig:mhd_local});
   This indicates that stars closely trace the underlying gravitational potential;
   \item in order to show any self-similarity, we have introduced the relative measure, $\zeta = d_\mathrm{MH}/r$, whose distribution resembles a log normal when using all the clusters and contours pairs; the parameter $\zeta$ could be used to study the relaxation state of a cluster;
   the maximum of this distribution is located at $\zeta \sim 0.1$, thus the typical $d_\mathrm{MH}$ is about $10 \%$ of the distance at which it is computed.
\end{enumerate}

The study of the spatial distribution of the ICL can be used to infer, in high detail, the distribution of the underlying dark matter in clusters of galaxies. Moreover, the average density profile of total matter can be extracted, and extrapolated up to the virial radius, only by measuring the slope of the stellar mass density profile and its normalisation close to the centre of the cluster. This is complementary to the study of \cite{Pillepich2018}, where only the total halo mass was given as function of the slope of the 3D stellar density profile, with larger uncertainty.

\section*{Acknowledgements}

We are very grateful to Ignacio Trujillo and Mireia Montes for supporting this work with useful ideas and discussions.
CDV acknowledges the support of the Spanish Ministry of Science, Innovation and Universities (MCIU) through grants RYC-2015-18078 and PGC2018-094975-B-C22.
YMB acknowledges funding from the EU Horizon 2020 research and innovation programme under Marie Sk{\l}odowska-Curie grant agreement 747645 (ClusterGal) and the Netherlands Organisation for Scientific Research (NWO) through VENI grant 639.041.751.
This work used the DiRAC@Durham facility managed by the Institute for Computational Cosmology on behalf of the STFC DiRAC HPC Facility (www.dirac.ac.uk). The equipment was funded by BEIS capital funding via STFC capital grants ST/K00042X/1, ST/P002293/1, ST/R002371/1 and ST/S002502/1, Durham University and STFC operations grant ST/R000832/1. DiRAC is part of the National e-Infrastructure.

\bibliographystyle{mnras}
\bibliography{ICL_refs}

\bsp
\label{lastpage}
\end{document}